\documentclass[aps,prc,reprint,floatfix,superscriptaddress]{revtex4-2}

\usepackage{graphicx}
\usepackage{amsmath}
\usepackage{amssymb}
\usepackage{bm}
\usepackage{xcolor}
\usepackage[colorlinks=true, breaklinks=true, linkcolor=blue, citecolor=blue, urlcolor=blue]{hyperref}

\newcommand{\Pspace}{\mathcal{P}}
\newcommand{\Qspace}{\mathcal{Q}}

\begin{document}

\title{Assessing continuum channel importance in
continuum-discretized coupled-channels via dynamic
polarization potential decomposition}

\author{Jin Lei}
\email[]{jinl@tongji.edu.cn}
\affiliation{School of Physics Science and Engineering, Tongji University, Shanghai 200092, China.}
\affiliation{Southern Center for Nuclear-Science Theory (SCNT), Institute of Modern Physics, Chinese Academy of Sciences, Huizhou 516000, Guangdong Province, China.}

\author{Hao Liu}
\affiliation{School of Physics Science and Engineering, Tongji University, Shanghai 200092, China.}

\date{\today}

\begin{abstract}
A recurring question in continuum-discretized coupled-channels
calculations is which continuum channels carry the breakup coupling,
both to interpret the reaction and to decide which channels a model
space can safely omit. The standard answer is bin deletion: remove a
channel, re-solve the coupled equations, and read the change in the
elastic $S$ matrix. We show that this cannot isolate an individual
channel's contribution. Deleting a channel forces the surviving model
space to reorganize, with the neighboring bins rerouting through the
off-diagonal Green's function, so the recorded change mixes the
channel's own effect with the readjustment of all the others. Working
instead from the channel-resolved Feshbach dynamic polarization
potential (DPP), whose full-coupling Green's function is a fixed
reference shared by every channel, we define an exclusion that removes
one channel while holding that reference fixed, returning its
contribution to the intact system. The two operations disagree
on which continuum bins matter most, for $d$+$^{58}$Ni at 21.6~MeV even
sending the intrinsically leading bin to last place under deletion. The
DPP further
separates each channel's action into a direct path, virtual excitation
and return, and a bridge path, relayed through neighboring bins, a split
deletion cannot make; it shows the bins act on the elastic channel
mainly as bridges, robustly across angular momentum, and that it is this
bridge coupling that reorganizes under deletion. A deletion-based
channel importance is therefore best read as the truncated calculation's
sensitivity to a channel's removal, not as the channel's intrinsic
coupling strength.
\end{abstract}

\maketitle


\section{Introduction}
\label{sec:intro}

In nuclear reactions involving weakly bound projectiles, coupling to
breakup continuum channels can profoundly alter elastic scattering and
reaction cross sections~\cite{Canto2015}. The continuum-discretized
coupled-channels (CDCC) method~\cite{Rawitscher1974,Austern1987,
Yahiro2012} handles this by discretizing the projectile continuum into
energy bins and solving the resulting coupled equations. A persistent
practical question is which continuum channels contribute most to the
breakup coupling effect, a question that is essential for
understanding breakup-modified elastic scattering, guiding model-space
truncation, and interpreting converged results.

In CDCC, each continuum bin is a square-integrable wave packet labeled
by the relative orbital angular momentum $l$ of the projectile
fragments and a narrow excitation-energy interval; together with the
projectile ground state, a chosen set of such bins spans the model
space in which the dynamics are evaluated. The projectile-target
couplings among the states of this space define the coupled radial
equations whose solution yields the elastic and breakup $S$ matrices
at each total angular momentum $J$, and any observable is converged by
enlarging the model space (raising the partial-wave cutoff
$l_{\max}$ and the bin-energy cutoff $E_{\max}$) until the result
stabilizes.

The standard approach has been bin deletion: remove a channel from the
coupled equations, re-solve, and measure the change in the elastic
$S$ matrix~\cite{Hirabayashi1991}. This technique has been widely applied
to compare resonant and nonresonant breakup
states~\cite{GomezCamacho2015,GomezCamacho2016,GomezCamacho2018,
Hirabayashi1991}, to quantify continuum-continuum coupling
effects~\cite{Canto2009}, to identify dominant breakup
components~\cite{Howell2005,Watanabe2015}, to map coupling form
factors~\cite{Nunes2004}, and to study model-space
convergence~\cite{Druet2012}. However, because deletion removes a
channel from the coupled set entirely, it also severs the bridge
pathways that run through that channel, and the surviving channels
propagate through a reorganized Green's function: the measured response
reflects not only the deleted channel's own contribution but also the
readjustment of the surviving basis. What it produces is thus a
property of the truncated calculation, not an intrinsic property of the
channel.

\begin{figure*}[t]
\centering
\includegraphics[width=\textwidth]{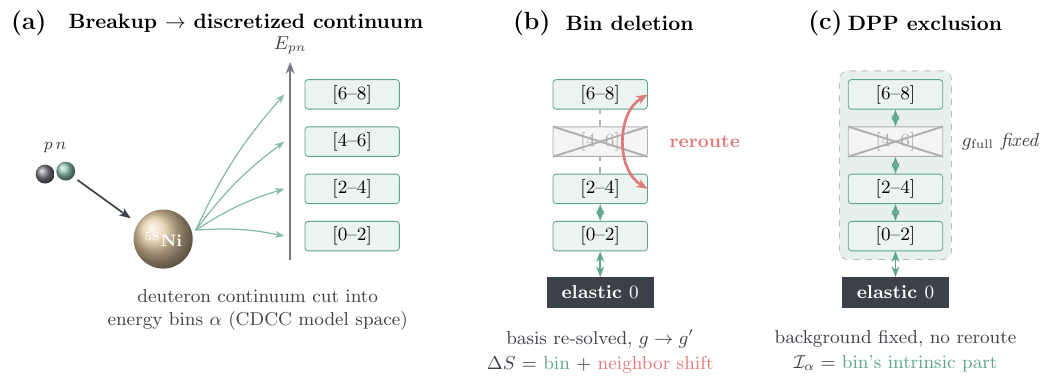}
\caption{Two ways to assess a continuum channel's importance in CDCC.
(a)~The projectile continuum is discretized into energy bins $\alpha$;
the bins, together with the projectile ground state, span the model
space coupled to the elastic channel. (b)~Bin deletion removes a
channel from the coupled set and re-solves, so the Green's function is
rebuilt ($g\rightarrow g'$, the reduced-space $g^{(-\alpha)}$ of
Sec.~\ref{sec:dpp}) and the surviving bins reroute through the
off-diagonal coupling; the recorded $\Delta S$ mixes the deleted bin's
own contribution with the readjustment of its neighbors. (c)~The DPP
exclusion removes the same channel while holding the full-coupling
Green's function $g_{\rm full}$ fixed as a shared reference, leaving the
other bins untouched and returning the channel's intrinsic contribution
$\mathcal{I}_\alpha$.}
\label{fig:schematic}
\end{figure*}

The Feshbach projection
formalism~\cite{Feshbach1958,Feshbach1962,FeshbachRMP1964} provides a
channel-importance measure that avoids this complication. The effect
of all eliminated channels on elastic scattering is encoded in the
dynamic polarization potential (DPP), a nonlocal effective
interaction that has been extracted from coupled-channel $S$ matrices
by inversion~\cite{Ioannides1986,Mackintosh2004}. The breakup-induced
DPP was first characterized by Sakuragi, Yahiro, and
Kamimura~\cite{Sakuragi1986}, who found it repulsive and weakly
absorptive, qualitatively unlike a collective-coupling DPP; later work
decomposed it by continuum subspace, computing the DPP from coupling to
resonant versus nonresonant states~\cite{GomezCamacho2015,
GomezCamacho2016,GomezCamacho2018}. The present work carries this DPP
decomposition down to the individual bin and, within each bin,
separates the direct pathway from the bridge pathway. The full-coupling,
channel-resolved DPP recently computed in our companion
paper~\cite{LiuLeiRen2026} enables a direct channel-by-channel
decomposition: the DPP is a bilinear sum over channel pairs, each term
linking the elastic channel out to one continuum channel and back in
from another through the full-coupling Green's function, so that every
term is labeled by a pair of channels. A channel can then be removed by
dropping the terms that carry its label, with that Green's function held
fixed, whereas deletion re-solves a reduced coupled system. Each channel
is thereby assessed against an
identical dynamical background, a measure of its contribution to the
intact system rather than of the truncated system's response to its
removal (Fig.~\ref{fig:schematic}). In this work we use this construction to show that deletion and the
intrinsic measure rank the continuum bins differently: at $J = 8$
deletion ranks the [0.32--0.39]~fm$^{-1}$ bin highest while the
intrinsic measure ranks [0.39--0.47]~fm$^{-1}$, a reordering we
trace, both algebraically and numerically, to the reorganization of
the surviving model space. The decomposition further resolves each
channel's contribution into a direct (self-coupling) part and a bridge
(interchannel coherence) part, a separation unique to the DPP route,
and reveals that the continuum bins act on the elastic channel
predominantly through bridges; this bridge dominance, robust across
angular momentum, is exactly the part of the dynamics that reorganizes
under deletion.

The argument unfolds in three steps. We first cast the
channel-importance question in terms of the channel-resolved DPP,
define the intrinsic importance measure, and split it into direct and
bridge parts (Sec.~\ref{sec:dpp}). We then apply it to $d+{}^{58}$Ni,
where the measure and standard bin deletion rank the bins differently,
a gap we trace to the reorganization of the surviving model space, and
where the direct/bridge decomposition shows the coupling is
bridge dominated (Sec.~\ref{sec:results}). We close by drawing out what
this means for channel-importance diagnostics in CDCC
(Sec.~\ref{sec:discussion}).


\section{Dynamic polarization potential and its decomposition}
\label{sec:dpp}

A continuum bin can act on elastic scattering in two physically
distinct ways: directly, by carrying the projectile into the bin and
back to its ground state, or as a bridge, by relaying the flux through
the coupling network of the other bins. The Feshbach
formalism~\cite{Feshbach1958,Feshbach1962} makes this separation exact.
Projecting onto the elastic channel ($\Pspace$) and onto the eliminated
continuum bins ($\Qspace = 1 - \Pspace$), the elastic-channel effective
Hamiltonian acquires the dynamic polarization potential
\begin{equation}
\Delta U(R,R') = \sum_{\gamma,\gamma'}
U_{0\gamma}(R)\, g_{\gamma\gamma'}(R,R')\, U_{\gamma' 0}(R'),
\label{eq:dpp}
\end{equation}
where $R$ is the projectile-target radial coordinate, $U_{0\gamma}$
couples the elastic channel to continuum bin $\gamma$ (labeled by
the orbital angular momentum $l$ and the excitation-energy interval), and
$g_{\gamma\gamma'}(R,R')$ are the channel matrix elements of the
full-coupling Green's function $G^{(+)}_J(R,R')$, which retains all
continuum-continuum couplings~\cite{LiuLeiRen2026,LiuLeiRen2026GF}.
Here $G^{(+)}_J = (E - H^{J}_{\Qspace\Qspace})^{-1}$ is the resolvent of the
$\Qspace$-space Hamiltonian at total angular momentum $J$, so
Eq.~(\ref{eq:dpp}) is the exact Feshbach
polarization potential for the elastic channel, with no weak-coupling or
perturbative approximation.
The off-diagonal elements $g_{\gamma\gamma'}$ ($\gamma \neq \gamma'$)
carry the coherence between eliminated bins that builds the bridge
paths, and are absent in weak-coupling
treatments~\cite{Canto2009}.

Grouping the double sum by channel participation decomposes the DPP
into direct and bridge contributions for each channel~$\alpha$:
\begin{align}
\Delta U^{(\alpha)} &= D_\alpha + B_\alpha, \label{eq:decomp}\\
D_\alpha &= U_{0\alpha}\,g_{\alpha\alpha}\,U_{\alpha 0}, \nonumber\\
B_\alpha &= \!\sum_{\gamma\neq\alpha}
\bigl(U_{0\alpha}\,g_{\alpha\gamma}\,U_{\gamma 0}
+ U_{0\gamma}\,g_{\gamma\alpha}\,U_{\alpha 0}\bigr), \nonumber
\end{align}
with $\Delta U = \sum_\alpha D_\alpha + \frac{1}{2}\sum_\alpha
B_\alpha$, the factor $\tfrac{1}{2}$ correcting the double counting of
each off-diagonal bridge pair, which appears once in $B_\alpha$ and
again in $B_\gamma$. The direct term $D_\alpha$ represents virtual excitation
elastic${}\to\alpha\to{}$elastic via the diagonal propagator
$g_{\alpha\alpha}$; the bridge term $B_\alpha$ captures multi-state
paths mediated by the off-diagonal $g_{\alpha\gamma'}$, so a bin with
weak direct coupling can still contribute through coherence with its
neighbors.

To weigh a single bin $\alpha$, we drop it from the sum in
Eq.~(\ref{eq:dpp}) while leaving $g_{\gamma\gamma'}$ untouched. This is
an algebraic operation on the already-computed DPP, not a fresh
coupled-channel solve: the Green's function is built once and reused, so
removing $\alpha$ changes nothing about how any other bin propagates. We
measure the effect on the elastic $S$ matrix through
\begin{equation}
\mathcal{I}_\alpha = \frac{|S_\mathrm{full} -
S_{\mathrm{no}\,\alpha}|}{|S_\mathrm{full}|},
\label{eq:importance}
\end{equation}
where $S_\mathrm{full}$ and $S_{\mathrm{no}\,\alpha}$ are the elastic
$S$-matrix elements at fixed total angular momentum $J$, and
$\Delta S_\alpha \equiv S_\mathrm{full} - S_{\mathrm{no}\,\alpha}$.
Because every bin is weighed against the same fixed continuum, the
importances $\mathcal{I}_\alpha$ are directly comparable from one bin to
the next; this is the intrinsic measure we use throughout. They are not
additive, since $S$ responds nonlinearly to $\Delta U$: what the
construction provides is not additivity but a common, unmoving
reference. We call $\mathcal{I}_\alpha$ \emph{intrinsic} in this precise
sense, that every channel is weighed against an identical, unmoving
dynamical background; it remains a property of the channel within the
fully coupled system, not of an isolated channel, which has no meaning
once the continuum is coupled.

This exclusion has a transparent meaning inside CDCC itself. The
$\Qspace$-space Hamiltonian, and hence $g$, depends only on the
continuum-continuum couplings and never on the elastic couplings
$U_{0\gamma}$, so setting $U_{0\alpha} = U_{\alpha 0} = 0$ in the
coupled-channels Hamiltonian and re-solving leaves $g$ pointwise
unchanged and returns the same $S_{\mathrm{no}\,\alpha}$. The operation
closes the elastic doorway of bin $\alpha$ while leaving it fully
embedded in the network of continuum-continuum couplings through the untouched
$U_{\gamma\alpha}$ ($\gamma \neq 0$): what it isolates is the channel's
effect with the rest of the network held fixed. A direct CDCC
re-solve with $U_{0\alpha} = U_{\alpha 0} = 0$ reproduces the fixed-$g$
exclusion importance to within $0.4$ percentage points across the
midcontinuum bins at $J = 8$, rising for the two highest bins near the
channel-closure threshold to the level of the reconstruction floor, the
numerical accuracy with which the DPP rebuilds the full elastic
$S$ matrix (Sec.~\ref{sec:results}). The equality is exact at the operator level, since $g$ is
independent of the elastic couplings $U_{0\alpha}$; the residual is
numerical, of the same origin as that reconstruction floor, not a
breakdown of the identity. Codes such as FRESCO~\cite{Thompson1988} specify couplings by
form factor and partition rather than by individual matrix elements, so
this re-solve, like the DPP construction itself, operates at the
matrix-element level; of the two the DPP route is the more economical,
since one full-coupling Green's-function evaluation assesses every
channel at once. The decomposition into direct and bridge terms,
Eq.~(\ref{eq:decomp}), is intrinsic to the DPP and cannot be recovered
from $S$ matrix differences alone: it names the physical pathway by
which each channel acts, one-step excitation versus coherence with its
neighbors.

Standard bin deletion does something different and more drastic. It
takes $\alpha$ out of the $\Qspace$ space altogether and solves the
coupled equations again in the smaller space that is left. The two
operations share one part and differ by one part, and the difference can
be written down exactly. This is the comparison the rest of the paper
turns on, so we set it out term by term. Exclusion, which drops
$\alpha$ from the sum in Eq.~(\ref{eq:dpp}) while holding $g$ fixed,
removes only the terms that contain $\alpha$,
\begin{equation}
\delta U_\alpha^{\rm excl} = \Delta U^{(\alpha)} = D_\alpha + B_\alpha,
\label{eq:excl}
\end{equation}
exactly the channel contribution $\Delta U^{(\alpha)}$ already written in
Eq.~(\ref{eq:decomp}). At fixed $g$ the change produced by excluding $\alpha$,
$\delta U_\alpha^{\rm excl}$, coincides with the contribution assigned to
$\alpha$ by the decomposition, $\Delta U^{(\alpha)}$; the exclusion
operation and the decomposition thus yield the same quantity. The factor
$\tfrac{1}{2}$ in $\Delta U = \sum_\alpha D_\alpha +
\tfrac{1}{2}\sum_\alpha B_\alpha$ does not enter the single-channel
exclusion. It corrects a double counting present only in the summation
over all channels: each off-diagonal bridge pair contributes to both
$B_\alpha$ and $B_\gamma$, so $\sum_\alpha B_\alpha$ enumerates every pair
twice. The exclusion of a single channel involves no such summation.
Because the bridge connecting $\alpha$ and $\gamma$ cannot exist in the
absence of $\alpha$, excluding $\alpha$ omits both of the terms in which
it appears, those carrying $g_{\alpha\gamma}$ and $g_{\gamma\alpha}$, from
the sum at once, so the full $B_\alpha$ is removed rather than
$\tfrac{1}{2}B_\alpha$. Deletion removes these same terms, but it does one
more thing. The reduced system is itself a coupled-channels problem, so
its elastic-channel effective potential has the same exact Feshbach form
as Eq.~(\ref{eq:dpp}), only built on the Green's function
$g^{(-\alpha)}$ of the one-bin-smaller continuum space: with the
continuum space reduced, the propagator itself changes, from $g$ to
$g^{(-\alpha)}$. The total change recorded by deletion is therefore
\begin{equation}
\begin{gathered}
\delta U_\alpha^{\rm del} = \delta U_\alpha^{\rm excl} + R_\alpha, \\
R_\alpha = \!\!\sum_{\gamma,\gamma' \neq \alpha}\!\!
U_{0\gamma}\bigl(g_{\gamma\gamma'} - g^{(-\alpha)}_{\gamma\gamma'}\bigr)
U_{\gamma' 0}.
\end{gathered}
\label{eq:reorg}
\end{equation}
Equation~(\ref{eq:reorg}) is an exact identity, not an approximation. The
extra piece $R_\alpha$, which we call the \emph{reorganization term}, is
built only from the surviving bins, since its sum runs over
$\gamma,\gamma' \neq \alpha$ and never touches $\alpha$ itself. It
measures how much the surviving bins' own contribution to the elastic
channel shifts once $\alpha$ is pulled out of the propagator $g$ they all
share, and it vanishes when $\alpha$ decouples from the surviving bins,
being generically nonzero otherwise. In short, exclusion records $\Delta U^{(\alpha)}$, the
contribution of bin $\alpha$ against a fixed background, while deletion
records $\Delta U^{(\alpha)} + R_\alpha$, that same contribution plus the
readjustment of every other bin to a background that has changed.

The reorganization term has a clear physical meaning. Deleting a bin
does more than erase that bin's own contribution: it forces every
surviving bin to propagate through a Green's function that has lost one
of its coupling partners, so the surviving bins shift as well. The magnitude of this shift can be
made explicit. Splitting the $\Qspace$-space
Hamiltonian into a block for $\alpha$ and a block for the survivors and
inverting, the survivors' Green's function comes out as the
reduced-space one, $g^{(-\alpha)}$, plus a correction built from the
flux that leaves the survivors, passes through $\alpha$, and comes back,
\begin{equation}
\begin{gathered}
g_{\gamma\gamma'} - g^{(-\alpha)}_{\gamma\gamma'} =
\bigl[g^{(-\alpha)}\,\Sigma_\alpha\,g^{(-\alpha)}\bigr]_{\gamma\gamma'}
+ \mathcal{O}(\Sigma_\alpha^2), \\
\Sigma_\alpha = V_{r\alpha}\,\frac{1}{E-H_{\alpha\alpha}}\,V_{\alpha r},
\end{gathered}
\label{eq:dyson}
\end{equation}
where $\Sigma_\alpha$ is exactly that round trip through $\alpha$, and
$V_{\alpha r}$ are the couplings of $\alpha$ to the survivors $r$.
These are the same couplings that build the bridge term $B_\alpha$
(through $g_{\alpha r} =
g_{\alpha\alpha}\,V_{\alpha r}\,g^{(-\alpha)}$), so, at leading order,
the reorganization is driven by the very matrix elements that make a
channel a bridge, and the bridge-dominated channels are the ones most
sensitive to deletion. This correspondence holds exactly at the first order shown;
the higher-order terms in Eq.~(\ref{eq:dyson}) involve longer paths
through $\alpha$, and the first-order term alone already accounts for the
reorganization-bridge correspondence found below. Its sign is not fixed: for complex, multichannel potentials
$R_\alpha$ can add to or subtract from the intrinsic shift, so deletion
neither systematically over- nor under-states a channel's contribution.
What deletion records, $\delta U_\alpha^{\rm excl} + R_\alpha$,
inseparably mixes the channel's own effect with the readjustment of all
the others. The two operations thus answer different questions: the
intrinsic contribution of a channel, and the sensitivity of the
truncated calculation to its removal.


\section{Results}
\label{sec:results}

We test the construction on $d$+$^{58}$Ni elastic scattering at
$E_\mathrm{lab} = 21.6$~MeV, asking whether bin deletion and the
intrinsic measure agree on which continuum bins carry the breakup
coupling. We run the comparison in the $l = 2$ sector, with the ground
state and the $l = 0$ and $l = 1$ couplings held fixed as the dynamical
background. The $l = 2$ quadrupole is the discriminating choice. The $l = 1$ dipole
is the longest-range, dominant breakup mode, but its strength is
concentrated in the low-energy bins just above threshold, a steep
hierarchy in which the leading bins are not in serious doubt; the
$l = 0$ monopole carries no off-diagonal Coulomb coupling and
contributes weakly. Neither sector, then, poses the question this work turns on:
whether the two diagnostics rank the bins differently.
The eight $l = 2$ bins instead carry comparable nuclear and
Coulomb-quadrupole strength spread from threshold to the open-channel
limit, where the
ranking is genuinely contested and the choice of diagnostic can matter. The two measures do not rank the $l = 2$ bins in the same
order: their rankings reorganize relative to each other, most strongly at
the grazing partial wave $J = 8$, on which we concentrate. The deuteron continuum is discretized into bins of
equal width in the $n$-$p$ relative momentum within each $l$ sector,
a standard CDCC prescription: six bins for $l = 0$ and $l = 1$ (relative
energy 0 to 12~MeV, $\Delta k = 0.087$~fm$^{-1}$, $k$ to $0.54$~fm$^{-1}$),
and, for $l = 2$, which carries the importance analysis, a ladder of
eight bins extended nearly to the open-channel limit
(relative energy 0 to 16~MeV, $\Delta k = 0.076$~fm$^{-1}$,
$k$ to $0.62$~fm$^{-1}$) so that the interpreted bins are free of truncation. The $n$-$p$ interaction is a central Gaussian of depth
72.15~MeV and range 1.484~fm, reproducing the 2.224~MeV deuteron binding
energy~\cite{Austern1987}; the nucleon-target interactions are Koning--Delaroche optical
potentials~\cite{KoningDelaroche2003} at half the deuteron laboratory
energy; the coupled equations are solved by the Lagrange-mesh $R$-matrix
method (200 basis functions, 60~fm matching radius), with the potentials
tabulated on a 0.05~fm radial grid. The DPP exclusion and the CDCC deletion use the
same Hamiltonian, bins, and numerical parameters and differ only in how
a channel is removed, so any disagreement between them reflects the
difference between the two questions, not between two calculations. As a
control on the decomposition, solving the one-channel elastic equation
with the bare (folded) potential plus the DPP of Eq.~(\ref{eq:dpp})
reproduces the full CDCC elastic
$S$ matrix to within $3.1\%$, $1.5\%$, and $0.8\%$ at $J = 6$, $8$, and
$10$, in each case below the interpreted single-bin importances at that
$J$ (the lowest two bins, which sit at this reconstruction floor, are
not interpreted). The reconstruction is tightest
for a compact continuum and loosens as the ladder is carried toward the
channel-closure threshold. In
its present form the channel-resolved DPP is evaluated over the open
channels only, the closed-channel Whittaker continuation used by standard
deletion has not yet been implemented, and its reconstruction degrades when the
highest bin is pushed against that closure threshold, which sets the
eight-bin ladder used here; this is a limitation of the construction, not
of the comparison: both diagnostics are evaluated on the identical
open-channel basis, so no part of their disagreement can come from seeing
different model spaces, and the inequivalence itself
[Eq.~(\ref{eq:reorg})] is projection algebra that holds on any basis; the
individual importances, like any CDCC quantity, do depend on the basis. The operator identity
$U_{\rm CDCC} = U_{\rm bare} + U_{\rm DPP}$, with $U_{\rm bare}$ being the bare
elastic (folded) potential and $U_{\rm DPP} = \Delta
U$ being the polarization potential of Eq.~(\ref{eq:dpp}), is established and
shown to be exact in the companion
work~\cite{LiuLeiRen2026,LiuLeiRen2026GF}, so this residual is of
numerical, not physical, origin. The agreement
confirms that the
channel-resolved decomposition faithfully represents the CDCC dynamics.

The breakup-induced DPP has so far been available only in local,
approximate form: as an angular-momentum-averaged local potential in the
founding work~\cite{Sakuragi1986} and the foundational CDCC
review~\cite{Austern1987}, and resolved by continuum subspace, resonant
versus nonresonant, in the local-equivalent
prescription~\cite{GomezCamacho2015,GomezCamacho2016,GomezCamacho2018}.
The exact, fully nonlocal DPP, the rigorous Feshbach polarization
potential evaluated with the full-coupling Green's function, was first
constructed in our companion work~\cite{LiuLeiRen2026,LiuLeiRen2026GF}.
What is new here is to resolve that exact DPP down to the individual
continuum bin and, within each bin, to separate the
direct pathway from the bridge pathway.

\begin{table}[b]
\caption{Single-bin importance $\mathcal{I}_\alpha$ (\%) for the
$l = 2$ continuum bins at $J = 8$. $\mathcal{I}^{\rm DPP}$ is the DPP
channel exclusion (net of direct and bridge); $\mathcal{I}^{D}$ and
$\mathcal{I}^{B}$ are the direct-only and bridge-only exclusions, which
do not sum to $\mathcal{I}^{\rm DPP}$ because $S$ depends nonlinearly on
the potential; and $\mathcal{I}^{\rm del}$ is the standard CDCC bin deletion.
Ranks of $\mathcal{I}^{\rm DPP}$ and $\mathcal{I}^{\rm del}$ (in
parentheses, 1 = largest, each column ranked independently) are taken
over the six interpreted bins; the two
near-threshold bins, {[0.02--0.09]} and {[0.09--0.17]}, lie at the DPP
reconstruction floor ($\lesssim 1.5\%$), are not physically interpreted,
and are excluded from the ranking. The bins span
the $l = 2$ continuum up to $k = 0.62$~fm$^{-1}$, just below the
channel-closure threshold. The two
orderings disagree sharply: the intrinsic measure ranks the [0.39--0.47]~fm$^{-1}$
bin first while deletion ranks it last.}
\label{tab:importance}
\begin{ruledtabular}
\begin{tabular}{lcccc}
Bin $k$ (fm$^{-1}$) & $\mathcal{I}^{\rm DPP}$ & $\mathcal{I}^{D}$ &
$\mathcal{I}^{B}$ & $\mathcal{I}^{\rm del}$ \\
\hline
{[0.02--0.09]} & 0.0 & 0.0 & 0.1 & 0.0 \\
{[0.09--0.17]} & 1.5 & 0.9 & 2.3 & 1.8 \\
{[0.17--0.24]} & 8.6\,(5) & 6.7 & 12.7 & 9.7\,(4) \\
{[0.24--0.32]} & 11.6\,(3) & 15.3 & 21.1 & 18.7\,(2) \\
{[0.32--0.39]} & 9.5\,(4) & 15.9 & 22.5 & 20.8\,(1) \\
{[0.39--0.47]} & 18.5\,(1) & 12.0 & 28.0 & 1.7\,(6) \\
{[0.47--0.55]} & 16.4\,(2) & 5.9 & 19.4 & 17.8\,(3) \\
{[0.55--0.62]} & 4.1\,(6) & 3.3 & 4.1 & 4.4\,(5) \\
\end{tabular}
\end{ruledtabular}
\end{table}

The intrinsic importance of the $l = 2$ bins (Table~\ref{tab:importance})
shows a clear hierarchy: it is negligible for the two near-threshold
bins ([0.02--0.09] and [0.09--0.17]~fm$^{-1}$, at the reconstruction
floor and not interpreted) and is largest in the upper continuum,
peaking at the [0.39--0.47]~fm$^{-1}$ bin ($18.5\%$) and then falling
toward the open-channel edge. The hierarchy is
not a simple function of bin energy, but it mirrors the strength of the
couplings themselves: the doorway form factors $U_{0\alpha}(R)$ are
concentrated at the nuclear surface and are strongest for the
midcontinuum bins, peaking at [0.39--0.47]~fm$^{-1}$, the same bin the
importance ranks first at $J = 8$ and $J = 10$. The importance peak also
migrates with $J$ (Table~\ref{tab:Jsweep}), from [0.24--0.32]~fm$^{-1}$
at $J = 6$ to [0.39--0.47]~fm$^{-1}$ at $J = 8$ and $J = 10$. The static
form factors cannot carry this migration, being independent of $J$ and,
if anything, shorter ranged for the higher bins; the migration is a
property of the propagation, of how the rising centrifugal barrier
reweights the contributions of the bins. The overall
$l = 2$ importance falls steeply with $J$, as the rising
barrier screens the surface region where the coupling is concentrated.

The direct/bridge decomposition [Eq.~(\ref{eq:decomp})] explains why a
single coupling-strength number is the wrong picture. In every
interpreted bin the bridge exclusion exceeds the direct one
($\mathcal{I}^{B} > \mathcal{I}^{D}$, Table~\ref{tab:importance}; at the
highest open bin the margin lies within the reconstruction floor), so
each bin acts on the elastic channel mainly by relaying flux through its
neighbors rather than by coupling back on its own; $\mathcal{I}^{D}$ and
$\mathcal{I}^{B}$ switch off, respectively, a bin's diagonal term and all
off-diagonal terms involving it, one at a time; they are separate
single-pathway removals, not
additive parts of $\mathcal{I}^{\rm DPP}$. The [0.32--0.39]~fm$^{-1}$ bin
makes the point sharply: its net importance
($\mathcal{I}^{\rm DPP} = 9.5\%$) is smaller than either single-pathway
value ($\mathcal{I}^{D} = 15.9\%$, $\mathcal{I}^{B} = 22.5\%$). The direct-only
and bridge-only removals shift $S$ by
$\Delta S^{D} = (-0.027, -0.047)$ and $\Delta S^{B} = (+0.049, +0.059)$,
two large, nearly antiparallel vectors in the complex plane (relative
phase $170^\circ$), while removing the channel altogether leaves only the
small net $\Delta S = (+0.027, +0.017)$. The modest net importance is
thus a coherent near-cancellation between strong direct and bridge
couplings acting in nearly opposite directions, not a weak underlying
coupling; because $S$ is nonlinear in $\Delta U$, these two single-pathway
shifts need not sum to the net.
That a channel acts on the elastic scattering mainly through this
back-and-forth, multi-step coupling rather than through a one-step
return is in keeping with earlier dynamic polarization potentials for
weakly bound systems, in which the elastic cross section is shaped by
the coupling back and forth between the bound and continuum
states~\cite{AndresGomezCamacho1999}.

\begin{figure}[t]
\centering
\includegraphics[width=\columnwidth]{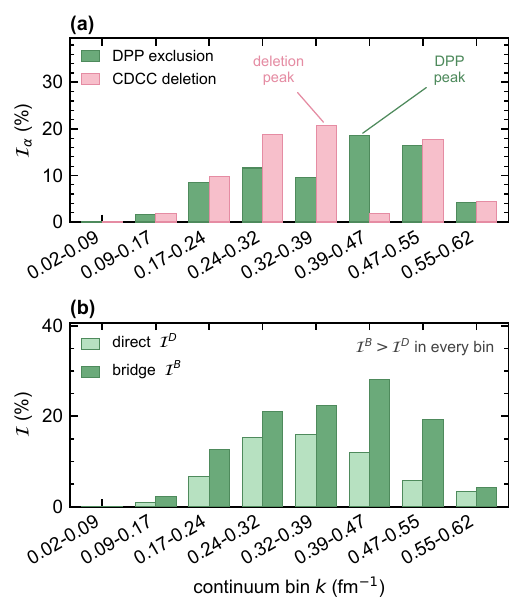}
\caption{Single-bin importance
$\mathcal{I}_\alpha = |\Delta S_\alpha|/|S_\mathrm{full}|$ for
$d$+$^{58}$Ni at 21.6~MeV, $J = 8$, $l = 2$. (a)~DPP channel exclusion
versus standard CDCC bin deletion: the two rank the bins differently,
deletion peaking at [0.32--0.39]~fm$^{-1}$ and the intrinsic measure at [0.39--0.47]~fm$^{-1}$. (b)~Direct-only ($\mathcal{I}^{D}$)
versus bridge-only ($\mathcal{I}^{B}$) exclusion: the bridge
contribution exceeds the direct one for all interpreted bins and does
so decisively across the midcontinuum, the
signature of bridge-dominated coupling.}
\label{fig:importance}
\end{figure}

Standard deletion, shown alongside the intrinsic measure in
Table~\ref{tab:importance} and Fig.~\ref{fig:importance}, gives a
different ordering: it ranks the
{[0.32--0.39]}~fm$^{-1}$ bin most important and [0.24--0.32]~fm$^{-1}$ second, whereas the
intrinsic measure ranks [0.39--0.47]~fm$^{-1}$ first and [0.47--0.55]~fm$^{-1}$ second. The
disagreement is not marginal: the bin the intrinsic measure ranks first,
{[0.39--0.47]}~fm$^{-1}$, deletion ranks last of the six interpreted bins, and the two orderings stay
inequivalent at $J = 6$ as well, agreeing on the leading bin only at $J = 10$
(Table~\ref{tab:Jsweep}); they differ most at $J = 8$. Nor is deletion a uniform rescaling of the
intrinsic measure: at $J = 8$ the ratio
$\mathcal{I}^{\rm DPP}/\mathcal{I}^{\rm del}$ runs from $0.5$ at
{[0.32--0.39]}~fm$^{-1}$ to above $10$ at [0.39--0.47]~fm$^{-1}$, where the bin the intrinsic
measure ranks first is nearly invisible to deletion, so deletion overstates some channels
and understates others.

This is the behavior predicted by Eq.~(\ref{eq:reorg}): deletion records
$\delta U_\alpha^{\rm excl} + R_\alpha$, and the reorganization
$R_\alpha$, built from the same bridge couplings, shifts each channel by
a different, sign-indefinite amount. The data bear this out. Define the
reorganization signature $|\Delta S_\alpha^{\rm del} - \Delta
S_\alpha^{\rm excl}|$, the complex gap between the deletion and exclusion
shifts. Across the interpreted $l = 2$ bins it peaks together with the
bridge importance $\mathcal{I}^{B}_\alpha$, at
the high-bridge [0.39--0.47]~fm$^{-1}$ bin; this is exactly the bin on
whose rank deletion and the intrinsic measure disagree most, the intrinsic
measure ranking it first and deletion last, so the disagreement is driven
by the bridge-built reorganization. Deletion thus reports how sensitive the truncated
calculation is to a channel's removal, not the channel's intrinsic
contribution.

\begin{figure}[t]
\centering
\includegraphics[width=\columnwidth]{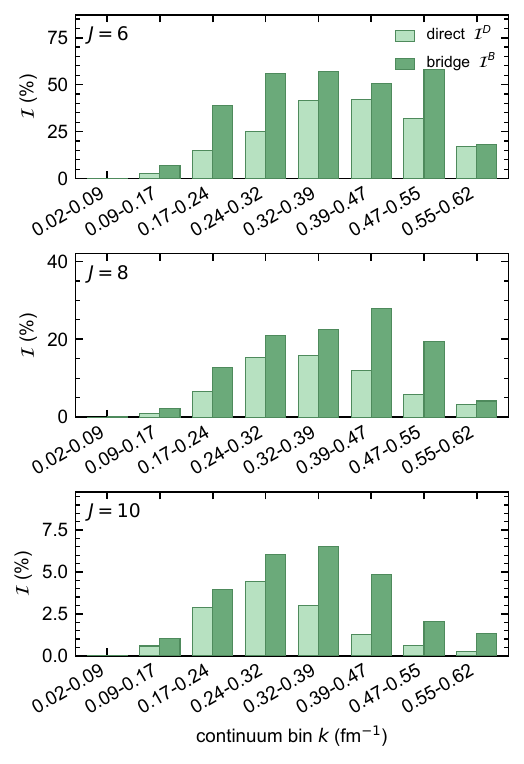}
\caption{Angular-momentum robustness of bridge dominance.
Direct-only ($\mathcal{I}^{D}$, light) and bridge-only
($\mathcal{I}^{B}$, dark) exclusions for the $l = 2$ bins at
$J = 6, 8$, and $10$. The bridge contribution exceeds the direct one in
all interpreted bins at every angular momentum and does so decisively
across the midcontinuum, so the bridge-dominated
character of the continuum coupling is not specific to $J = 8$.}
\label{fig:J_robustness}
\end{figure}

The same picture holds away from $J = 8$. At $J = 6$ and $J = 10$
(Fig.~\ref{fig:J_robustness}) the bridge dominance persists:
$\mathcal{I}^{B} > \mathcal{I}^{D}$ for every interpreted $l = 2$ bin at
every angular momentum, with margins that exceed the reconstruction
floor in every midcontinuum bin, by factors from about $1.5$ to more
than $10$; only at the highest open bin at $J = 6$
and $8$ does the margin fall within the floor, where the ordering is
consistent with, rather than established by, the calculation. The
conclusion that the continuum couples to
the elastic channel mainly through interchannel coherence therefore
rests on no single bin or angular momentum. The deletion-versus-intrinsic ranking, by
contrast, is angular-momentum dependent (Table~\ref{tab:Jsweep}): the two
orderings differ at $J = 6$ and $J = 8$, most strongly at $J = 8$, and
agree on the leading bin only at $J = 10$. In
every case the per-bin ratio
$\mathcal{I}^{\rm DPP}/\mathcal{I}^{\rm del}$ spreads on both sides of
unity. Deletion therefore provides neither the
intrinsic ranking nor a fixed rescaling of the intrinsic magnitude at
any $J$, consistent with the reorganization term $R_\alpha$ of
Eq.~(\ref{eq:reorg}) being channel dependent and sign indefinite.

\begin{table}[t]
\caption{Single-bin $l = 2$ importance (\%) from DPP exclusion
($\mathcal{I}^{\rm DPP}$) and CDCC deletion ($\mathcal{I}^{\rm del}$)
at $J = 6, 8$, and $10$, with ranks in parentheses (1 = largest, each column
ranked independently). Ranks are over the same six
interpreted bins at every $J$; the two near-threshold bins
({[0.02--0.09]}, {[0.09--0.17]}), which sit at the reconstruction floor
($\lesssim 1.5\%$) at $J = 8$, are excluded throughout for uniformity. The bins extend up to $k = 0.62$~fm$^{-1}$, just below the channel-closure threshold.
The two orderings are inequivalent at $J = 6$ and $J = 8$, agreeing on the
leading bin only at $J = 10$, with the strongest reordering at $J = 8$.}
\label{tab:Jsweep}
\begin{ruledtabular}
\begin{tabular}{lcccccc}
 & \multicolumn{2}{c}{$J=6$} & \multicolumn{2}{c}{$J=8$} &
   \multicolumn{2}{c}{$J=10$}\\
\cline{2-3}\cline{4-5}\cline{6-7}
Bin $k$ (fm$^{-1}$) & DPP & del & DPP & del & DPP & del \\
\hline
{[0.02--0.09]} & 0.0 & 0.1 & 0.0 & 0.0 & 0.0 & 0.0 \\
{[0.09--0.17]} & 3.9 & 2.5 & 1.5 & 1.8 & 0.6 & 0.7 \\
{[0.17--0.24]} & 24.8\,(4) & 9.8\,(6) & 8.6\,(5) & 9.7\,(4) & 2.2\,(5) & 3.2\,(5) \\
{[0.24--0.32]} & 40.9\,(1) & 16.7\,(5) & 11.6\,(3) & 18.7\,(2) & 2.6\,(3) & 4.2\,(3) \\
{[0.32--0.39]} & 39.1\,(2) & 36.3\,(1) & 9.5\,(4) & 20.8\,(1) & 4.1\,(2) & 3.5\,(4) \\
{[0.39--0.47]} & 19.3\,(5) & 20.0\,(3) & 18.5\,(1) & 1.7\,(6) & 4.4\,(1) & 7.7\,(1) \\
{[0.47--0.55]} & 38.8\,(3) & 28.6\,(2) & 16.4\,(2) & 17.8\,(3) & 2.4\,(4) & 4.5\,(2) \\
{[0.55--0.62]} & 17.5\,(6) & 19.8\,(4) & 4.1\,(6) & 4.4\,(5) & 1.3\,(6) & 1.2\,(6) \\
\end{tabular}
\end{ruledtabular}
\end{table}

A channel-importance analysis is only meaningful if the continuum ladder
is itself converged, free of an artificial accumulation of flux in the
highest bin when that bin has no higher partner to couple to. The $l = 2$
ladder used throughout approaches the open-channel limit: eight bins out to
$k = 0.62$~fm$^{-1}$, just below the channel-closure threshold
($E_{\rm c.m.} - B_d \approx 18.7$~MeV, $k \approx 0.67$~fm$^{-1}$,
where $B_d = 2.224$~MeV is the deuteron binding energy paid to
dissociate it) at
which the $d$+$^{58}$Ni breakup channels close. The elastic $S$ matrix is
converged on this basis: enlarging the $l = 2$ ladder from six to seven
to eight bins (which also refines the momentum grid) shifts it by only
$2.8\%$ and then $1.1\%$, and the per-bin
importance turns over at high $k$, the highest open bin
($k = 0.55$--$0.62$~fm$^{-1}$) being small in both diagnostics
($\mathcal{I}^{\rm DPP} = 4.1\%$, $\mathcal{I}^{\rm del} = 4.4\%$;
Table~\ref{tab:importance} and Fig.~\ref{fig:importance}). There is thus
no flux pile-up at the truncation edge; the importance is largest in the
mid-$k$ region, where the doorway form factors are strongest, and falls
off smoothly toward the open-channel
boundary. Extending the $l = 0$ and $l = 1$ ladders to the same limit
alongside $l = 2$ leaves the $l = 2$ per-bin importance essentially
unchanged: no single-bin importance moves by more than 2 percentage
points, the deletion ordering is identical in the two bases, and the
DPP ordering changes only by an interchange of its fourth- and
fifth-ranked bins, so the profile is a property of the
continuum dynamics, not of an unbalanced truncation.


\section{Discussion and outlook}
\label{sec:discussion}

The picture that emerges is coherent and practical. Bin deletion, the
standard tool for judging continuum channel importance, answers a
well-defined question, but not the one it is usually assumed to answer.
Removing a channel sets off a reorganization of the surviving model
space [Eq.~(\ref{eq:reorg})], in which the neighboring bins, coupled to
the removed channel through the same matrix elements that build the
bridge term, reroute around it. What deletion records is therefore an
inseparable mixture of the channel's own contribution and this
reorganization, and it ranks the bins differently from the intrinsic
measure, the per-bin importance ratio at $J = 8$ running from $0.5$ to
above $10$. Deletion-based channel
importances in the
literature~\cite{GomezCamacho2015,GomezCamacho2016,Canto2009} are, on
this reading, best interpreted as a measure of how
sensitive a truncated calculation is to a channel's removal, not as the
channel's intrinsic contribution to the breakup coupling. The
complementarity runs both ways: that removal sensitivity is itself the
relevant quantity for the practical question of which channels a finite
model space may safely omit, so for model-space truncation the deletion
ranking is the appropriate diagnostic, whereas the intrinsic measure,
which holds the background fixed, ranks channels by physical
contribution and is not a guide to truncation efficiency.

Two results follow. The first is conceptual: the inequivalence of DPP
exclusion and coupled-channel deletion follows from the Feshbach
projection algebra [Eqs.~(\ref{eq:reorg}) and (\ref{eq:dyson})] and
holds for any projected coupled-channels system, with the reorganization
term $R_\alpha$ governed by the bridge couplings and, in general,
sign indefinite. The $l = 2$ results shown here are therefore
representative; by the same algebra the inequivalence is expected to
apply to the $l = 0$ and $l = 1$ sectors as well, though we do not
quantify it there. The second is physical: the direct/bridge
decomposition, available only from the DPP, shows that the continuum
bins couple to the elastic channel predominantly as bridges, by relaying
flux through neighboring bins rather than by coupling back on their own,
a feature robust across angular momentum. Both require the
channel-resolved DPP, hence the full-coupling Green's function
$G^{(+)}_J(R,R')$ of the companion work~\cite{LiuLeiRen2026GF}, which
standard CDCC codes such as FRESCO~\cite{Thompson1988} do not generate.
Where that machinery is unavailable, a deletion-based importance from a
standard CDCC code should still be read as a ranking of truncation
sensitivity, not of intrinsic coupling strength; the direct versus
bridge character cannot be recovered from deletion at all, since
deleting a channel removes both contributions together while
reorganizing the survivors in one inseparable step.

The $d$+$^{58}$Ni case studied here is a representative, well-controlled
test. We expect, though do not demonstrate here, the effects it isolates
to be strongest in halo systems such
as $^{6}$He, $^{11}$Li, and $^{11}$Be, where the spatially extended
projectile is likely to enhance the off-diagonal Green's function
couplings:
there the continuum coupling should be even more thoroughly
bridge dominated, the reorganization under deletion correspondingly
larger, and the gap between what a channel intrinsically contributes and
what its removal appears to cost at its widest. These are precisely the
reactions in which the choice of which continuum channels to retain most
strongly shapes the predicted observables.

\begin{acknowledgments}
This work was supported by the National Natural Science Foundation of
China (Grants No.~12475132 and No.~12535009) and the Fundamental Research
Funds for the Central Universities. Large language models were used during
manuscript preparation for drafting and for cross-checking derivations,
numerical tables, and text-data consistency; every number and statement
so checked was verified by the authors against the coupled-channel
calculations and the raw $S$-matrix outputs.
\end{acknowledgments}

\section*{Data availability}
There are no publicly available research data or software supporting
this manuscript. Requests for further information or data should be
sent to the authors.

\bibliography{references}

\end{document}